\documentclass[conference]{IEEEtran}
\IEEEoverridecommandlockouts
\usepackage{cite}
\usepackage{amsmath,amssymb,amsfonts}
\usepackage{graphicx}
\usepackage{textcomp}
\usepackage{xcolor}

\usepackage{algorithm}
\usepackage{algpseudocode}
\usepackage{subfigure}
\usepackage{multirow}
\usepackage{diagbox}

\def\BibTeX{{\rm B\kern-.05em{\sc i\kern-.025em b}\kern-.08em
    T\kern-.1667em\lower.7ex\hbox{E}\kern-.125emX}}
\begin{document}

\renewcommand{\algorithmicrequire}{ \textbf{Input:}}      
\renewcommand{\algorithmicensure}{ \textbf{Output:}}     

\title{Active Link Obfuscation to Thwart Link-flooding Attacks for Internet of Things\\
}

\author{
	\IEEEauthorblockN{1\textsuperscript{st} Xuyang Ding{*}}
	\IEEEauthorblockA{\textit{School of Computer Science and Engineering} \\
	\textit{University of Electronic Science and Technology of China}\\
	Chengdu 611731, China \\
	dxy@uestc.edu.cn}
\and
	\IEEEauthorblockN{2\textsuperscript{nd} Feng Xiao}
	\IEEEauthorblockA{\textit{School of Cyber Science and Engineering} \\
	\textit{Wuhan University}\\
	Wuhan 430072, China \\
	f3i@whu.edu.cn}
\and
	\IEEEauthorblockN{3\textsuperscript{rd} Man Zhou{*}}
	\IEEEauthorblockA{\textit{School of Cyber Science and Engineering} \\
	\textit{Wuhan University}\\
	Wuhan 430072, China \\
	zhouman@whu.edu.cn}
\and
	\IEEEauthorblockN{4\textsuperscript{th} Zhibo Wang}
	\IEEEauthorblockA{\textit{School of Cyber Science and Engineering} \\
	\textit{Wuhan University}\\
	Wuhan 430072, China \\
	zbwang@whu.edu.cn}
}

\maketitle

\begin{abstract}
The DDoS attack is a serious threat to Internet of Things (IoT). As a new class of DDoS attack, Link-flooding attack (LFA) disrupts connectivity between legitimate IoT devices and target servers by flooding only a small number of links. In this paper, we propose an active LFA mitigation mechanism, called Linkbait, that is a proactive and preventive defense to throttle LFA for IoT. We propose a link obfuscation algorithm in Linkbait that selectively reroutes probing flows to hide target links from adversaries and mislead them to identify bait links as target links. To block attack traffic and further reduce the impact in IoT, we propose a compromised IoT devices detection algorithm that extracts unique traffic patterns of LFA for IoT and leverages support vector machine (SVM) to identify attack traffic. We evaluate the performance of Linkbait by using both real-world experiments and large-scale simulations. The experimental results demonstrate the effectiveness of Linkbait.
\end{abstract}


\section{Introduction}\label{sec:introduction}
\IEEEPARstart{T}{he} rapid growth of Internet of Things (IoT) devices has boosted various smart applications. For example, IoT devices equipped with camera,  motion sensor and microphone utilize visible light, movement and voice signals to transmit data among IoT devices~\cite{b1,b2}, perceive the outside world~\cite{b3,b4} and perform IoT devices authentication~\cite{b5}, etc. However, there are also various attacks~\cite{b6,b7,b8} against the IoT system. Botnet-driven distributed denial-of-service (DDoS) attack~\cite{b9}, is one of the most serious threats to the IoT system~\cite{b10,b11,b12,b13}.
For example, Mirai malware took advantage of compromised IoT devices to break down Internet connectivity of America by flooding links in October 2016~\cite{b14}.

A new type of sophisticated link-flooding based DDoS attacks has been proposed recently~\cite{b15,b16}, which utilizes distributed botnets to deplete the bandwidth of key network links and disrupt the network connectivity of the victims. LFA has been employed to construct real-world attacks~\cite{b17}, which can be easily captured by traditional defense mechanisms.

Recently, several defenses have been proposed to detect and mitigate it~\cite{b18,b19,b20,b21,b22,b23,b24,b25,b26,b27,b28,b29,b30}. However, these mechanisms mainly take effects after networks have been congested.
Therefore, it is necessary to design a preventive mechanism that captures adversaries behaviors earlier and defend against the attacks in advance so that we can preventively mitigate LFA for IoT without target links being congested by adversaries.

In this paper, we propose an active link obfuscation mechanism, called Linkbait, to actively mitigate LFA for IoT by constructing a fake linkmap to cheat adversaries. Inspired by Moving Target Defense (MTD)~\cite{b31}, we propose a link obfuscation algorithm to generate fake linkmap by selectively rerouting probing flows to obfuscate link information in the topology. Thus, target links are hidden from adversaries and meanwhile bait links that do not in the paths to the victims will be treated as target links by adversaries.In particular, Linkbait leverages a bait link construction strategy and randomly select flow rerouting policy to reduce the probability of bait links being congested by the attack. Furthermore, in order to completely rule out the attack traffic generated by adversaries, we develop a compromised IoT devices detection algorithm in Linkbait that extracts unique traffic patterns from the traffic generated by LFA for IoT. It leverages support vector machine (SVM) to accurately distinguish compromised IoT devices from legitimate IoT devices and block the attack traffic generated by compromised IoT devices.
There are three major challenges in defeating LFA for IoT: \emph{link obfuscation}, \emph{LFA resistance for bait links} and \emph{compromised devices detection}. We propose Linkbait to solve these challenges.

The remainder of this paper is organized as follows. We introduce the system model and the background of LFA in Section \ref{sec:background}. We present the design of Linkbait in Section \ref{sec:Linkbait} and the discussion in Section \ref{sec:Security_analysis}. We then evaluate the performance of Linkbait in Section \ref{sec:evaluation}. Finally, we conclude the paper in Section \ref{sec:conclusion}.

\section{Related work}

In this section, we briefly discuss the state-of-the-art of LFA detection and mitigation.

\noindent{\textbf{Attack detection:}} MoveNet~\cite{b26} employs virtual networks to offer constant, dynamic and threat-aware reallocation of critical network resources to deceive attacker's knowledge about critical network resources.
By updating routers into SDN-enable nodes and installing corresponding measurement indicators in these nodes in advance, Woodpecker~\cite{b27} quickly locates the congestion link in LFA by combining path analysis with hop-by-hop probing. In order to detect and mitigate LFA, \cite{b29} also leverages features of SDN, such as programmability, network-wide view, and flow traceability, to get the flowpaths by flow analysis, monitoring target links, rerouting traffic and blocking malicious traffic.

\noindent{\textbf{Mitigation of flooding traffic:}}
Aydeger et al. proposed a SDN based model leveraging TE dynamically to reroute traffic on the suspected target links as long as it is congested~\cite{b25}.
Kang et al. designed a SDN based system, called SPIFFY~\cite{b24}, that leverages temporary bandwidth extension to identify flooding traffic during LFA happens. Ma et al. proposed two novel mechanisms, called incentivized-optimal-routing and rerouting-on-demand~\cite{b30}, to stimulate the cooperation between ASes to mitigate LFA.

\section{Preliminaries}\label{sec:background}
In this section, we first introduce the link-flooding attack (LFA), and then present the system model.

\subsection{Link-flooding Attack}\label{sec:Link_flooding_attack}
Link-flooding attack targets links in the core of the network and creates a large number of attack flows crossing the targeted links to flood and virtually disconnect them. There are mainly two kinds of them: The first is the Coremelt attack~\cite{b15}. It utilizes bots to send attack traffic to other bots. And the second is the Crossfire attack~\cite{b16}, which coordinates bots to send legitimate-looking low-rate traffic to servers.
In particular, adversaries tend to choose the latter when they target at large networks. Hence, we mainly focus on the Crossfire attack in this paper and we employ LFA to denote such kind of attacks. LFA can be described as the following two steps: link information gathering and flooding.

\noindent{\textbf{Link information gathering}:} To launch LFA, the adversary will use all his bots to query link information towards as many servers in target area as possible.
The link information gathered by the adversary is called \emph{linkmap}. We call any host which performs such link information querying for legitimate or malicious purpose as a \emph{link-prober}. The links which can be attacked by enough bots are chosen as \emph{target links}.

\noindent{\textbf{Flooding}:} The adversary manipulates a large number of bots to persistently send TCP-like flows to congest the target links by consuming their bandwidth. Note that a rational adversary will cautiously manipulate his bots in a reasonable rate to avoid being detected by the rate-based detecting mechanisms.

\subsection{System model}

In this paper, as shown in Fig. \ref{fig:system_overview}, we focus on networks with two edges. IoT devices outside the network can access the servers in the target area via the ingress routers.
An adversary need to manipulate a large number of IoT devices to obtain the linkmap of the network by sending probing flows to the network. He figures out the target links between the ingress/egress routers, and then launches LFA to the target links in the target area.

\section{Linkbait design}\label{sec:Linkbait}

\begin{figure}[!t]
	\centering
	\includegraphics[width=0.85\columnwidth,height=0.6\columnwidth]{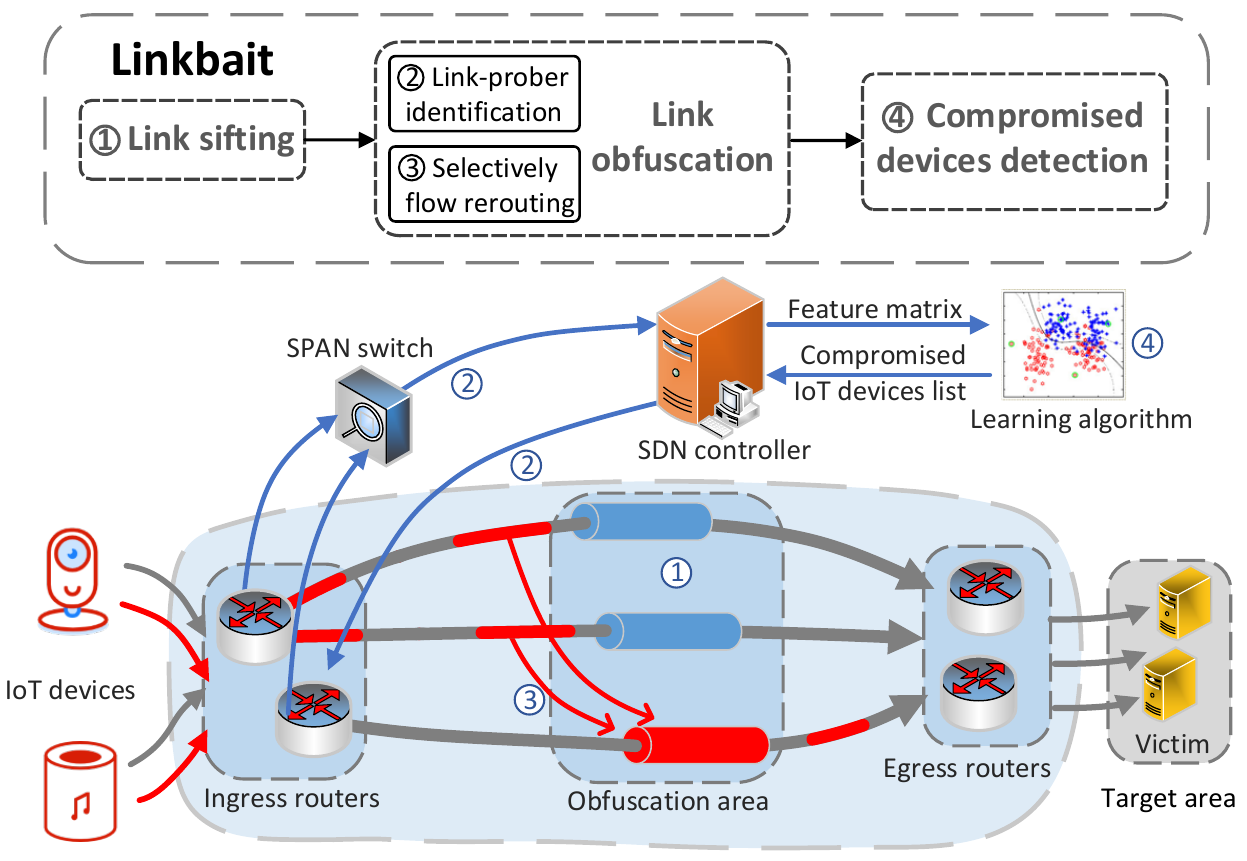}
	\caption{System overview with Linkbait.}
	\label{fig:system_overview}
\end{figure}

In this paper, we propose Linkbait to preventively mitigate LFA for IoT with active link obfuscation. The key idea of Linkbait is to provide an obfuscated linkmap to the adversary by imposing differential policies on probing flows, and mislead attacks from compromised IoT devices to the faked target links while hiding the true target links. Linkbait consists of three components, as shown in Fig. \ref{fig:system_overview}: \emph{link sifting}, \emph{link obfuscation} and \emph{compromised devices detection}.
\subsection{Link Sifting}

Link sifting aims to figure out potential target links of the network and select appropriate links to fake target links. We call the faked target link as \emph{bait links}. Link sifting has two phases: link analysis and link grouping.

\subsubsection{Link Analysis}
We propose a method to obtain the whole network information, called \emph{looking glass tracing} (LG tracing), which leverages existing network diagnostic tools to collect link information. There are many public available servers maintained by other ISPs, which the ISP can utilize to obtain the complete link information. We call these servers as LG servers and they can be remotely accessed for querying routing information. Therefore, with LG servers, LG tracing leverages existing network diagnostic tools to collect link information of the ISP~\cite{b33}.

Flow density represents bandwidth utilization for links. According to the universal power-law property of flow density distribution~\cite{b16}, the more flows can be created through one link, the higher flow density it has. Thus, we estimate the flow density of a link by calculating the number of LG servers going through the link.

\subsubsection{Link Grouping}
After gathering the link information, we can compute the flow density of each link and then figure out which links are most likely to be flooded. In this paper, we use the algorithm in~\cite{b16} to figure out the true target links.

In order to hide the true target links from the adversary, we select some links to fake target links and reroute the probing flows of compromised devices to these faked links. Hence, the adversary will obtain a fake linkmap and misjudge the bait links as the target links.
It is difficult to ensure that linkmap obfuscation on these links can affect enough flows in the network due to the limited link resources. To solve this problem, we design a link grouping algorithm that can cover as many flows as possible in the network when minimize the ``cost'' of adding links into bait links.

\noindent{\textbf{Problem Formulation:}}
We formulate the link grouping problem as a weighted set cover problem. Let $f_i$ denotes each single flow sending from individual IoT devices to the target network, and the total set of flows in the IoT is denoted by $ F = \{f_1, f_2, . . . , f_\varphi\}$. For each link, there are several flows going through it. Let $T(l_i) = \{f_{i_1},f_{i_2},\cdots\}$ be the set of flows going through the link $l_i$. Suppose there are $N$ flows, so the flows going through all the links are denoted by $L= \{T(l_1), T(l_2), . . . , T(l_N)\}$. It is obviously that $\cup_{i=1}^N T(l_i) = F$

In the weighted set cover problem. This nonnegative weight function $w : L \to \mathbb{R}$ is defined to reflect the cost of each link. A bait link with less cost is supposed to obfuscate more flows than others while their consumptions of link resources are the same. Let $b_i$ denote a bait link and $T(b_i)$ is the set of flows going through $b_i$. Our objective is to find a set of bait links which can minimize the total cost while covering as many flows as possible. With this objective, the link grouping problem is formulated as follows.
\begin{equation}
\begin{split}
\arg\min_B  & \sum_{b_i \in B} w(b_i) \\\text{s.t.}\ \ \ \  &\bigcup_{b_i \in B} T(b_i)=F
\end{split}
\end{equation}
where the weight of $b_i$, $w(b_i)$, can be characterized by two factors: the number of links in $b_i$, and the flow density of $b_i$. Note that the flow density of $b_i$, denoted by $\rho_{b_i}$, is the total number of flows in $T(b_i)$.

We first characterize the influence of $\rho_{b_i}$ to $w(b_i)$. Suppose there are $M$ IoT devices communicating with the target area. Let $F(h_i)$ denote the set of flows between servers in the target area and a IoT device $h_i$ outside the target area. Therefore, we have
\begin{equation}\label{Eq:overallTrafic}
\bigcup\limits_{i=1}^M F(h_i) = F = \bigcup\limits_{j=1}^N T(l_j)
\end{equation}

Generally speaking, linkmap obfuscation can be formulated as a \emph{Bernoulli experiment}: for any IoT device, $p_a$  approaches the overall proportion of its flows obfuscated in the network, which can be illustrated by \ref{Eq:possibility}. Hence, higher flow density of a bait link $\rho_{b_i}$ corresponds to a higher $p_a$. As a result, increasing flow density of bait links leads to a larger amount of compromised IoT devices obfuscated by Linkbait.
\begin{equation}\label{Eq:possibility}
p_a \approx  \sum_{b_i\in B} \rho_{b_i}/\|F\|
\end{equation}

\begin{algorithm}[!th]
	\caption{Greedy Link Grouping.}
	\label{alg:background}
	\hrule
	\begin{algorithmic}[1]
		\vspace{.1cm}
		\Require ~\\
		Total flows $F$;\\
		Flows grouped by links $L$;\\
		Bait link coverage threshold $\tau$;
		\Ensure ~\\
		Bait link set  $B$;
		\State B $\gets \emptyset $\ \ \ \ \% {initiate all sets}
		\State $F^{'} \gets F$
		\Repeat
		\State $l \gets argmax_{X\in L}|X\cap F|/w(X)$
		\State $B \gets B\cup l, L \gets L \verb|\|\{l\}$, and  $F^{'} \gets F^{'}\verb|\|L$
		\Until{$\|B\|/\|F\|>\tau$}\ \ \ \ \% {add proper links into B}
	\end{algorithmic}
	\hrule
\end{algorithm}

\noindent{\textbf{Grouping algorithm:}}
Since the weight set cover problem is a well-known NP-hard problem \cite{b34}, the formulated link grouping problem is a NP-hard problem. In this paper, we propose a greedy link grouping algorithm to solve the problem.

As mentioned above, a bait link should increase its flow density $\rho_{b_i}$ while maintain a small $n_{b_i}$. Hence, we construct bait links according to two principles. First, the links of a bait link can be chosen from several normal links which support a certain amount of traffic instead of links with very low flow density so that we can increase $\rho_{b_i}$. Second, in order to reduce  $n_{b_i}$, partial flows to true target links should also be rerouted to bait links. Since true target links usually have high $\rho_{b_i}$, redirecting their flows is an effective way to increase $\rho_{b_i}$ of bait links while keeping $n_{b_i}$ small at the same time. That is, Linkbait tries to find the links that best match the two principles (i.e., the least $w$) and combines them as bait links, until flows in bait links have a satisfying coverage to the total flows in the IoT. The description is stated in Algorithm \ref{alg:background}.

\subsection{Linkmap Obfuscation}

Linkmap obfuscation is proposed to provide a fake linkmap to the adversary and use several bait links to fake target links. It is proposed with two steps: link-prober identification and selectively flow rerouting.
\subsubsection{Link-prober Identification} This step aims to identify all the link-probers and label their probing flows for the rerouting purpose. In this paper, we leverage the SDN controller to analyze the mirrored traffic instead of performing this work on the ingress routers. The probing flows generated by traceroute have two unique features: repeated invalid destination port and different TTL from the same source, which can help us to distinguish the probing flows from other TCP-like flows.

Compromised IoT devices that send a sequence of packets containing different TTL and invalid dest ports, to collect link information toward the target area. With these two unique features, we can distinguish the probing flows from other flows, and the IoT devices creating probing flows will be identified as link-probers. After identifying the link-probers, flow tables are installed on the ingress routers to label probing flows of link-probers in a real-time manner.

\subsubsection{Selectively Flow Rerouting} In order to provide a fake linkmap to the adversary, we propose a selectively flow rerouting policy to reroute probing flows accordingly. The basic idea is to reduce probing flows to the target links and increase probing flows to the bait links, so that the bait links will be misjudged as target links by the adversary.

\begin{figure}[htbp]
\begin{minipage}[t]{0.45\linewidth}
	\centering
	\includegraphics[width=1\columnwidth,height=0.6\columnwidth]{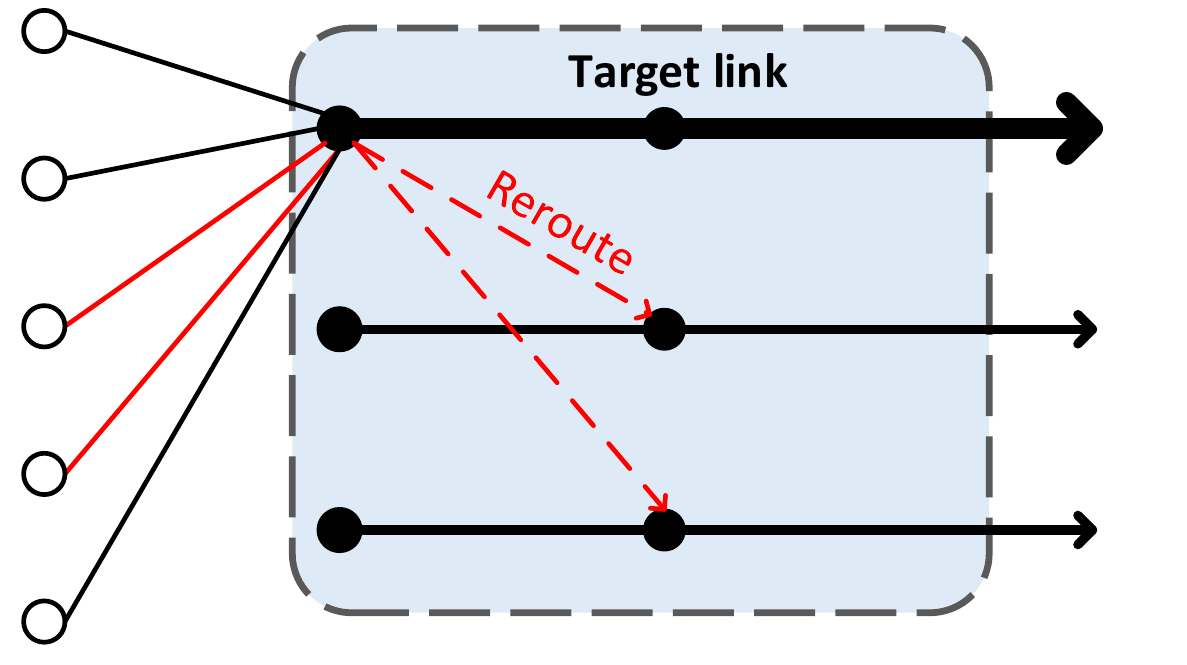}
	\caption{Rerouting policy for probing flows to a target link.}
	\label{fig:target_link}
\end{minipage}
\qquad
\begin{minipage}[t]{0.45\linewidth}
	\centering
	\includegraphics[width=1\columnwidth,height=0.6\columnwidth]{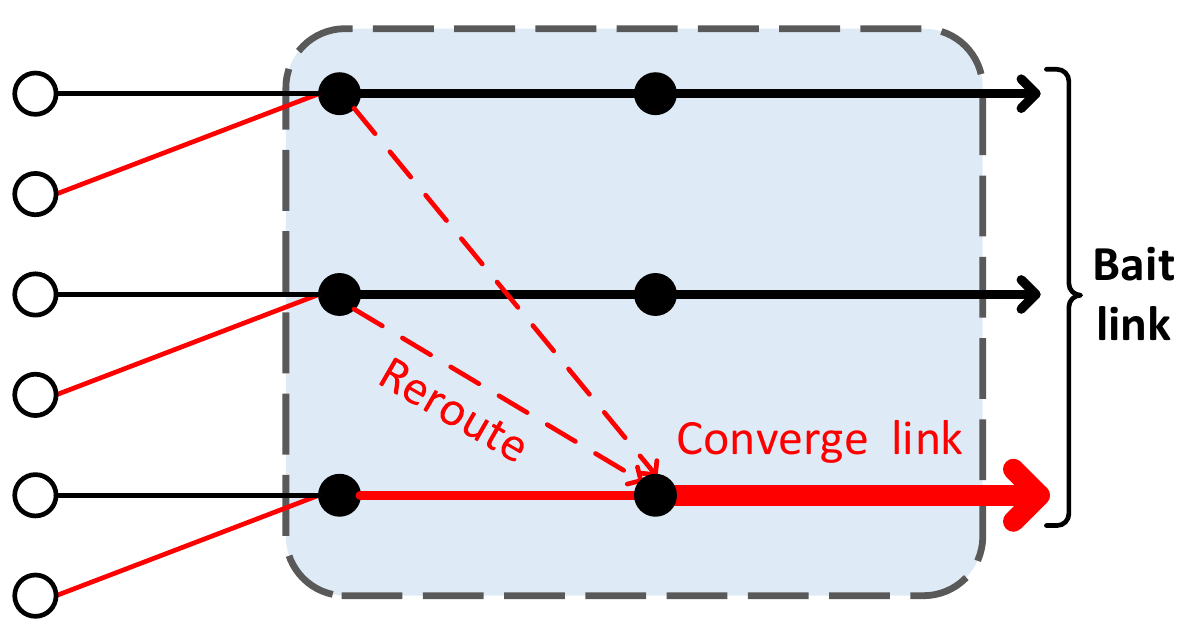}
	\caption{Rerouting policy for probing flows to a bait link.}
	\label{fig:sham_link}
\end{minipage}
\end{figure}

\noindent{\textbf{Rerouting policy for probing flows to the target link}:} In order to hide the target links from the adversary, we associate a set of links to each target link, which are called \emph{branch links}.  The branch links are chosen from the links that are close to the corresponding target links, which should have small communication latency with the target link. As shown in Fig. \ref{fig:target_link}, black lines which are entering the target link are legitimate flows whereas the red one are labeled probing flows. For each probing flow to the target link, it will be randomly rerouted to one of the branch links of the target link. This random rerouting policy reduces the flow density to each target link, and makes the link information changes dynamically to link-probers so they cannot figure out the true target links.

\noindent{\textbf{Rerouting policy for probing flows to the bait link}:} In order to fake a target link, for the links in a bait link, the one with the largest bandwidth will be selected as the converge link. As shown in Fig. \ref{fig:sham_link}, black lines entering the bait link are legitimate, and the red one are labeled probing flows. For any probing flow to the links of a bait link, it will be rerouted to the coverage link of the bait link. This will increase the flow density of the coverage link and mislead the judgement of the adversary.
\begin{figure}
	\centering
	\includegraphics[width=0.75\columnwidth,height=0.45\columnwidth]{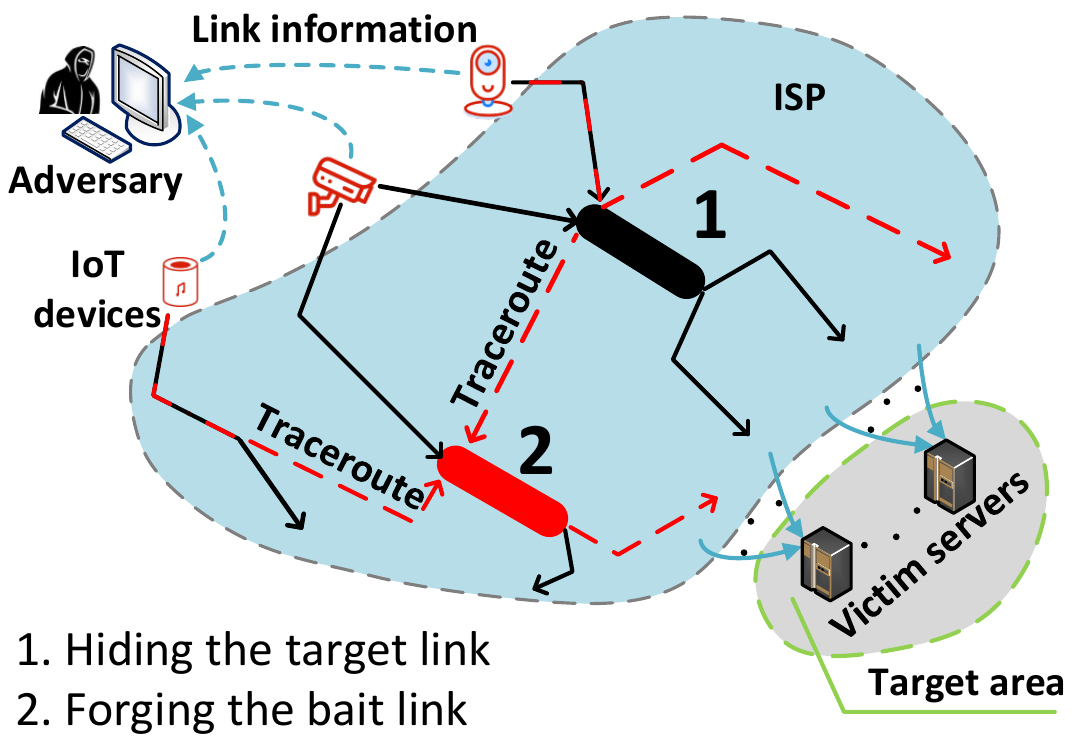}
	\caption{Illustration of selectively rerouting of probing flows.}
	\label{fig:Selectively_flow_rerouting}
\end{figure}

Fig \ref{fig:Selectively_flow_rerouting} illustrates how the probing flows from the compromised IoT devices of the adversary to the target area are selectively rerouted. According to the rerouting policy, only the probing flows (denoted by the red dashed) through the target link will be randomly distributed to its branch links, and the probing flows through the links of a bait link will be converged to its converge link, while the TCP-like flows (denoted by black lines) still go through their original links without any rerouting. Moreover, when the adversary launch attacks to the bait links with TCP-like flows, these flows will not be rerouted and still go through their original links. Hence, neither the true target links nor the bait links will be congested by the adversary, so that LFA becomes useless.

\subsection{Compromised devices detection}

In this paper, we propose to leverage the unique patterns during the linkmap construction phase and the flooding phase to accurately distinguish compromised devices from legitimate IoT devices. In particular, we monitor the long-term traceroute traffic and also continuously monitor the short-term flooding traffic with a sliding window. By combining these features, we leverage a supervised learning algorithm to accurately distinguish compromised devices from legitimate devices.

\subsubsection{Feature extraction}

We extract two features corresponding to the unique traffic patterns of the compromised devices during the flooding phase and the linkmap construction phase. The first feature is called flooding matrix, which represents the short-term flooding traffic patterns of a link-prober. The second feature is called traceroute matrix, which represents the long-term traffic patterns for linkmap construction of a link-prober.

\noindent{\textbf{Flooding Matrix (FM)}:} FM represents the flooding behaviors of an IoT device which accesses links during the flooding phase. However, it is unknown to us when the adversary will launch the flooding attacks. To solve this problem, we use a sliding window to continuously detect the flooding behaviors.

A sliding window consists of $n$ intervals and the sliding windows moves with one interval each time. Let $I_i$ denote the $i$th interval of a sliding window. We use $fs_{ij}$ to represent the traffic, denoted by the number of bytes, going through $link_i$ during $I_j$. The FM for a IoT device $i$ during a sliding window can be represented as follows.As the sliding windows moves, we can obtain a lot of FMs for each IoT device.
\begin{equation}\label{Eq:matrix1}
FM_{device_{i}}=\bordermatrix{%
	& I_1       & I_2     &\cdots     &I_n\cr
	link_1    & fs_{11}         &  fs_{12}        &\cdots     &  fs_{1n} \cr
	link_2    &  fs_{21}          &  fs_{22}        &\cdots     &  fs_{2n} \cr
	\vdots & \vdots    &\vdots   &\cdots     &\vdots\cr
	link_m    &  fs_{m1}          &  fs_{m2}        &\cdots     & fs_{mn}
}
\end{equation}

\noindent{\textbf{Traceroute Matrix (TM)}:} TM represents the traceroute behaviors of a link-prober during a detection period $DT$ whereas $DT$ is divided into multiple subperiods. Let $T_i$ denote the $i$th subperiod of $DT$. For each link-prober, we use $ft_{ij}$ to represent the traceroute frequency of the link-prober towards $link_i$ during $T_j$. Suppose there are $m$ links and $n$ subperiods. Thus, the TM for an IoT device $i$ is represented as follows.
\begin{equation}\label{Eq:matrix1}
TM_{device_{i}}=\bordermatrix{%
	& T_1       & T_2     &\cdots     &T_n\cr
	link_1    & ft_{11}         &  ft_{12}        &\cdots     &  ft_{1n} \cr
	link_2    &  ft_{21}          &  ft_{22}        &\cdots     &  ft_{2n} \cr
	\vdots & \vdots    &\vdots   &\cdots     &\vdots\cr
	link_m    &  ft_{m1}          &  ft_{m2}        &\cdots     & ft_{mn}
}
\end{equation}

Therefore, each link-prober has a TM to represent the traffic pattern during the detection period of $DT$. The value of $DT$ depends on how much time spent by the adversary to construct the linkmap, which varies from an adversary to another. We set $DT$ to be 5 days since adversaries usually take less than 5 days to obtain the linkmap.

The FM feature collects flow information on each bait link of all IoT devices in each time interval in the flooding period. The TM feature represents the traceroute behaviors on each bait link of all link-probers in the link information gathering period.
\subsubsection{Classification} With the extracted features, we then leverage a supervised classification algorithm to distinguish compromised devices from legitimate IoT devices. Each link-prober has one TM and many FMs while other IoT devices only have FMs. In particular, we combine all FMs together to form the joint-FM for each IoT device. In our experiments, we collect a ground truth dataset where each sample has a label to indicate the corresponding IoT devices is a compromised device or a legitimate user. In particular, a linear multiclass Support Vector Machine (SVM) classifier implemented by libSVM3~\cite{b37} is employed for accurate classification.

\section{Discussion}\label{sec:Security_analysis}

\subsection{Linkbait's Impact on adversaries}\label{sec:encounter}

The adversary aims at disjointing connections between the target area and Internet, so it tends to congest as many links as possible. Suppose each compromised device has an upstream bandwidth $U$.

\noindent{\textbf{LFA without Linkbait}:} To saturate a target link with bandwidth $B$, the adversary utilizes $N_{p}= B/U$ compromised devices whose flows can go through the target link. Using $N_{p}$ compromised devices guarantees a robust congestion even if there is no legitimate flows in the link. Let us denote the amount of these unused compromised devices as  $N_{un}$ ($N_{un}\ll N_{p}$). Thus, the number $N_b$ of compromised devices which the adversary need to finish the attack to a target area including $n$ target links is calculated as \begin{equation}
N_b =  n \cdot N_{p} + N_{un}
\end{equation}

\noindent{\textbf{LFA with Linkbait}:}
Suppose a bait link consists of $M$ normal links. When the adversary created TCP-like flows to congest the bait link, the flows will be distributed to $M$ links for each bait link. Let $\alpha_i B$ denote the bandwidth of the $i$ link of a bait link. In order to congest the bait link, the adversary is supposed to use $N_l$ compromised devices which is calculated as follows.
\begin{equation}
N_l = \frac {\alpha_1 B + \alpha_2 B + \cdots + \alpha_M B}{U} = \sum_{i=1}^M \alpha_i N_{p}
\end{equation}

That is, the number of compromised devices required to flood a bait link is $\sum_{i=1}^M \alpha_i N_{p}$. We can observe that if $\alpha_i=1$, the attack cost of the adversary has been forced to increase from $N_{p}$ to $M N_{p}$. Moreover, the more number of links to form a bait link, the higher the attack cost is required for the adversary.

\subsection{Link-prober identification is versatile to various probers}
In Linkbait, we identify a IoT device as a link-prober if it repeatedly creates flows to reach every hop of a link. The identification is feasible because a link-prober can only fetch informations of one hop in a link every time he requests if he wants to obtain link information towards the target area.

The adversary uses network diagnostic tools to collect link information of the network. The most significant characteristic of these tools is that it probes only one hop every time. Since a link-prober has no idea where his packet will be directed, he must query hops in the link repeatedly using tools like traceroute. Hence, the probing flows reveal the same feature no matter which network diagnostic tools are used.

\section{Performance Evaluation}\label{sec:evaluation}
In this section, we evaluate the performance of Linkbait by using both real-world experiments and large-scale simulations. In particular, we first implement link sifting in real-world networks. Then we implement a prototype on a real SDN testbed. Finally, we simulate large-scale LFA with large-scale simulations.

\subsection{Internet-scale Link Sifting} \label{sec:internet_sift}

We implement link sifting in real-world Internet to see whether we can find enough bait links to attract the attention of the adversary. We choose five places as the target areas in Internet. We leverage 126 LG servers provided by Telia~\cite{b38} and Cogentco~\cite{b39} to launch LG tracing.
\subsubsection{Basic link information}
The basic link information of the five ASes are collected using LG tracing, and the detailed information for each AS is shown in Table \ref{tab:basic}. The \emph{Avg Hop Num} represents the average number of hops in all paths towards each AS. The \emph{Non-identical Path} represents the number of different links from all LG servers to each AS. The flow density of a link is estimated by the times it appears in the results of traceroute of all the LG servers. The \emph{Coverage of Top-15 flow-density links} is the fraction of the number of LG servers which can access the top 15 highest flow density links to the total number of LG servers. It can be used to estimate the flow distribution of the network.
As seen in Table I, the coverage for AS1, AS3 and AS4 reach 100\% because almost all paths from different LG servers to each AS share the same link at the end servers, making the coverage 100\%. The network size can be seen from the non-identical paths. We observe that AS4 has the least number of paths whereas AS2 has the most, which indicates that AS4 is the smallest network whereas AS2 is the largest.
\begin{table}[!t]
	\caption{Basic link information for 5 ASes.}\label{tab:basic}
	\centering
	{\scriptsize
		\begin{tabular}{|c||c|c|c|}
			\hline
			\multirow{2}{*}{\backslashbox{\textbf{Area}}{\textbf{Info}} }   & \multirow{2}{*}{\textbf{Avg Hop Num } }    &\textbf{Non-identical}       &\textbf{Coverage of Top15}\\
			& &\textbf{Paths}         &\textbf{flow-density links}
			\\ \hline
			\hline
			
			\textbf{AS1}    & 12       & 603     & 1.0      \\ \hline
			\textbf{AS2}    & 12       & 804     & 0.84     \\ \hline
			\textbf{AS3}    & 16       & 497     & 1.0      \\ \hline
			\textbf{AS4}    & 14       & 353     & 1.0      \\ \hline
			\textbf{AS5}    & 12       & 792     & 0.81     \\ \hline
	\end{tabular}}
\end{table}

\subsubsection{Traceroute time cost}
\begin{figure}[!t]
	\centering
	\includegraphics[width=0.8\columnwidth,height=0.4\columnwidth]{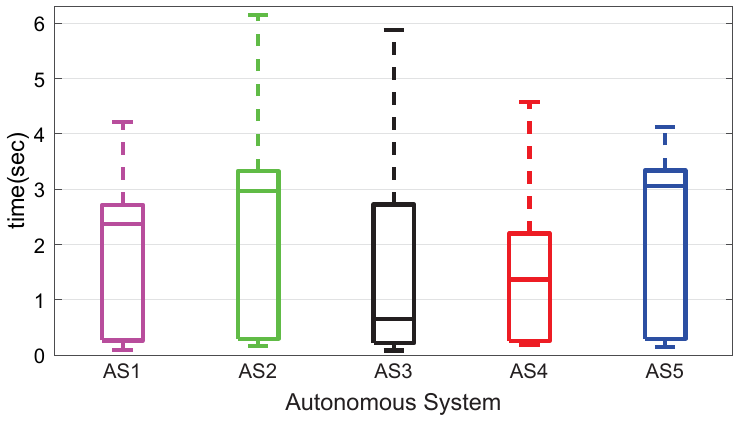}
	\caption{Time for traceroute.}
	\label{fig:tracetime}
\end{figure}

To investigate the relationship between link latency and the scale of ASes, we show the time spent of every single hop for each AS during traceroute in Fig. \ref{fig:tracetime}. We can see that the longest time spent for one hop is 6.14s due to network latency, whereas the lowest is 0.076s. The average time spent for discovering one node of all 5 ASes is 1.72s. Combining with Table \ref{tab:basic}, we observe that networks with more number of non-identical paths have a longer average time cost on every hop.

\subsubsection{Link coverage}
\emph{Link coverage} here is estimated by the fraction of the number of LG servers which can be obfuscate by bait links to the total number of LG servers. To be brief, link coverage denotes ratio of probing flows that will go through bait links. Hence, the higher the link coverage, the better obfuscation the network.

\begin{figure}[!t]
	\centering
	\includegraphics[width=0.8\columnwidth,height=0.4\columnwidth]{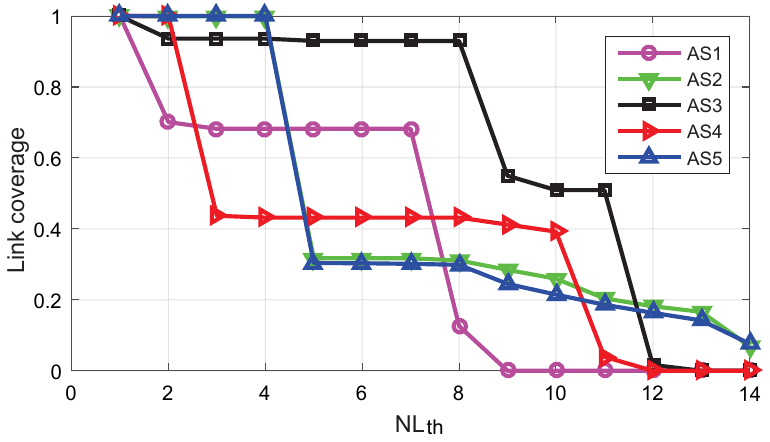}
	\caption{Link coverage for five target areas.}
	\label{fig:slc}
\end{figure}

In our experiments, we measure the link coverage for the 5 ASes with our grouping algorithm. Since the minimum number of links to form a bait link $NL_{th}$ demonstrates resistibility of Linkbait against flood, we change $NL_{th}$ while maintaining the least cost with our grouping algorithm.  Note that we cannot obtain the actual bandwidth or the flow density of links, $\rho_{b_i}$, in ISP, so we use the fraction of the number of LG servers that can travel through the link to estimate.
Fig. \ref{fig:slc} shows the link coverage against the variation of $NL_{th}$. We can see that the link coverage decreases as $NL_{th}$ increases, which is because less links in the network are chosen to form bait links. However, almost for all ASes, the link coverage can reach 70\% when $NL_{th} \leq 4$. The only exception is AS4 which mainly because its small-scale network has limited number of links for sifting. Based on these observations, we argue that link sifting can realize a satisfying linkmap obfuscation with an appropriate $NL_{th}$ that makes Linkbait stable enough.

\subsection{Evaluation using Real Testbed}

We then implement Linkbait in a real testbed to evaluate its performance. In particular, we focus on the rerouting latency introduced by Linkbait since our mechanism should not affect legitimate IoT devices or be perceived by the adversary. Therefore, the lower latency introduced by Linkbait, the better link obfuscation to the adversaries.

\subsubsection{Prototype Implementation}
We employ software defined network (SDN) to implement the DSCP based network prototype over physical nodes and links provided by Cloudlab~\cite{b40}. Cloudlab provides $2\times10$Gbps network interfaces to every node via SDN. We implement Linkbait on the \emph{Floodlight}~\cite{b41} controller. We use \emph{OVS}~\cite{b42} to virtualize layer-2 switches which support OpenFlow 1.3~\cite{b43} to perform selectively rerouting. In addition, we leverage \emph{iPerf} to emulate the legitimate TCP-like flows in the network.

We build a experimental network with two edges in the prototype, whose structure is similar to a simplified ISP network. A bait link with three parallel links is deployed between the edges. Among the three link, there is a converge link $L_c$ for link obfuscation, and there is another link $L_o$ through which both link-probers and legitimate users communicate with the target area. In addition, there is a link $L_l$ which only contains legitimates flows.
\subsubsection{Rerouting Latency}
The Rerouting latency is an important metric for evaluating the performance of Linkbait. Firstly, the network jitter caused by rerouting should not disturb legitimate IoT devices. Secondely, since latency may produce a deviation for the result of link-probers (traceroute), we should reduce this jitter in case the adversary perceives that the linkmap has been faked.

\begin{figure}
	\centering
	\includegraphics[width=0.7\columnwidth,height=0.4\columnwidth]{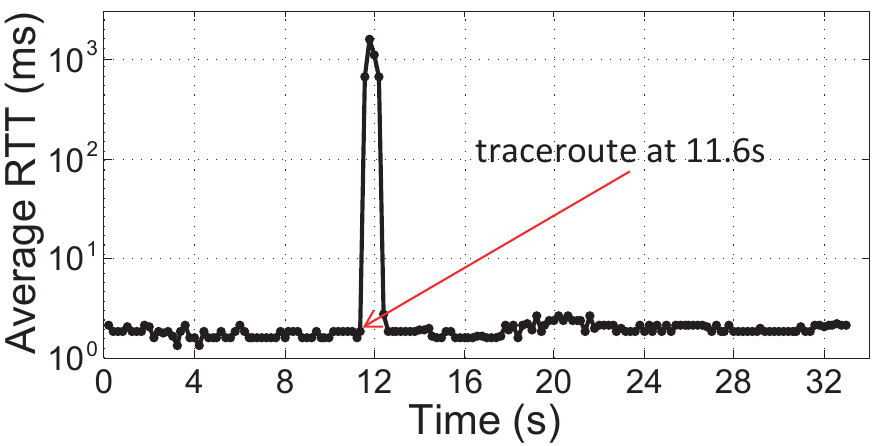}
	\caption{Real-time RTT change for legitimate devices.}
	\label{fig:RTT_change}
	\vspace{-3mm}
\end{figure}

\noindent{\textbf{Rerouting Impact on legitimate devices}:} To investigate the impact on legitimate IoT devices, we measure Round-Trip Time (RTT) during three different stages in Linkbait. Fig. \ref{fig:RTT_change} illustrates the RTT change for legitimate devices. The link $L_o$ supports all link-probers' traffic before 11.6s. Our mechanism identifies and reacts immediately after a link-prober performs traceroute at 11.6s. The response time includes the time for link-prober identification and the time for pushing corresponding labeling flow table to Openflow-enabled switches. As shown in Fig. \ref{fig:RTT_change}, legitimate IoT devices who communicate with the target area via $L_o$ experience a temporary block. However, RTT quickly returns to a normal value when Linkbait handles traceroute of the link-prober. 

\begin{figure}
	\centering
	\includegraphics[width=0.7\columnwidth,height=0.40\columnwidth]{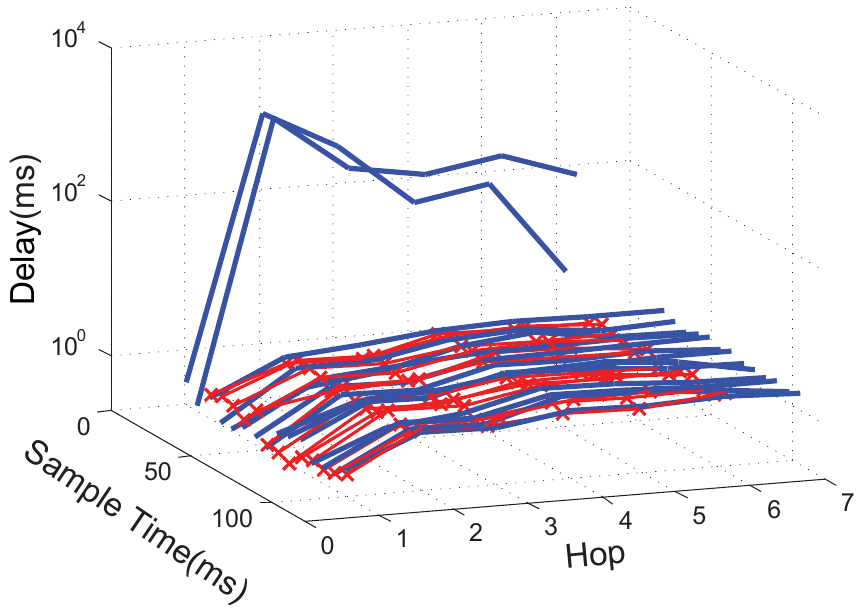}
	\caption{The timing information of traceroute time cost in a probing flow rerouting period.}
	\label{fig:Traceroute_change}
	\vspace{-3mm}
\end{figure}

\noindent{\textbf{Rerouting Impact on Link-probers}:} To investigate the impact on link-probers, we record the result of traceroute during rerouting policy takes effect. When traceroute runs, traceroute outputs the list of traversed routers in a simple text format, together with the timing information. With the list of routers, a link-prober restores a link. In addition, the link-prober can observe whether every hop works fine according to its traceroute delay. An abnormal delay may alert the adversary, which leads to a failed obfuscation.

Fig. \ref{fig:Traceroute_change} shows the timing information for traceroute command with and without our rerouting policy. We create probing flows every 7 ms into the network. Each blue line shows the average router response time for each hop of a probing flow with the rerouting policy. Each red line shows the average router response time for each hop of a probing flow without rerouting policy, which is considered as a baseline.

From Fig. \ref{fig:Traceroute_change},  we observe that probing flows experience an inevitable temporary block around the second hop when Linkbait starts to reroute its flow to $L_c$. This is because the rerouting operation occurs at that hop. We also observe that the latency of the following hops drops quickly, which indicates that our rerouting policy only blocks several hops rather than the whole link. Hence, Linkbait achieves a robust flow rerouting from $L_o$ to $L_c$. We also observe that time spent for querying every hop with the rerouting policy only slightly differs from that of baselines. The average time spent for every hop is 1.32ms whereas that of baselines is 1.24ms. As a consequence, Linkbait seamlessly obfuscates a linkmap before adversaries obtain the real one.

\subsection{Evaluation using Large-scale Simulation}\label{sec:simulation}
\begin{figure}[!t]
	\centering
	\subfigure[Compromised devices detection rate]{
		\includegraphics[width=0.48\columnwidth,height=0.4\columnwidth]{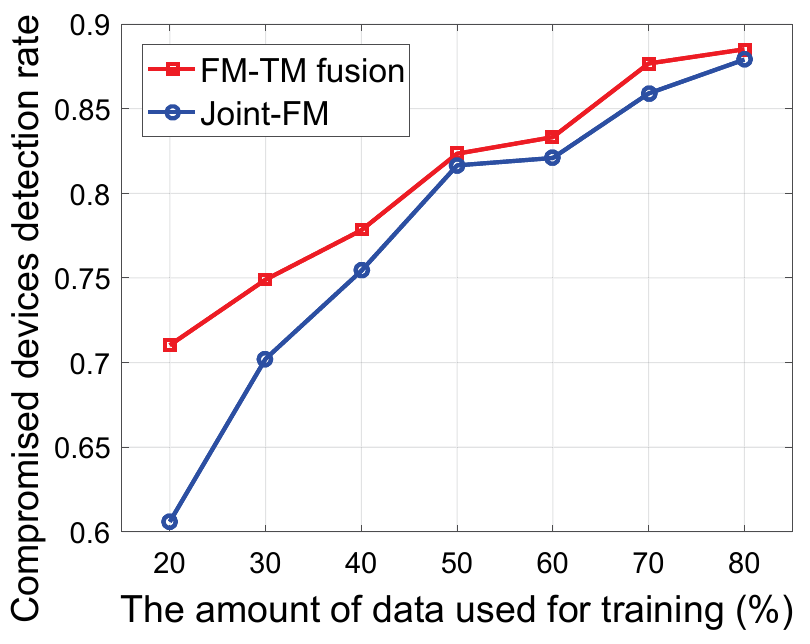}}
	\subfigure[False-positive rate]{
		\includegraphics[width=0.48\columnwidth,,height=0.4\columnwidth]{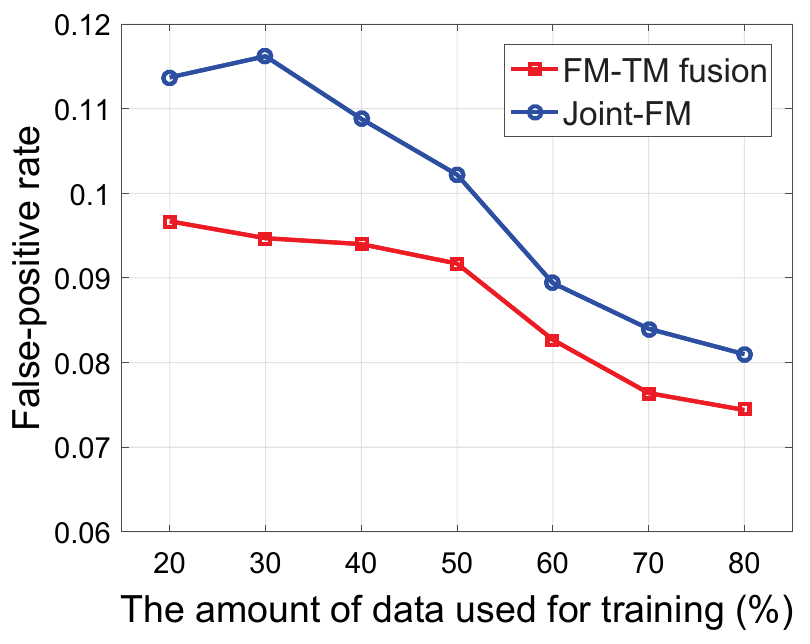}}
	\caption{The impact of the percentage of data used for training on the performance of compromised devices detection.}
	\label{fig:Detection_accuracy_trainrate}
\end{figure}

We use the real link information collected from the real-world Internet in Section \ref{sec:internet_sift} in the experiments of simulations. We deploy 100 compromised devices, 190 legitimate IoT devices and 20 servers in the target area. The legitimate IoT devices send packets to servers at different rates in order to get services from them. The compromised devices send flooding flows to launch LFA to the servers.

In Linkbait, we extract the FM and TM features and use SVM to distinguish compromised devices from legitimate IoT devices. The evaluation mainly focus on (1) \emph{compromised devices detection rate} and (2) \emph{false-positive rate}. Let $TP$ denote the number of correctly identified compromised devices, $TN$ denote the number of correctly identified legitimate IoT devices, $FP$ denote the number of legitimate IoT devices wrongly identified as compromised devices, $FN$ denote the number of compromised devices wrongly identified as legitimate IoT devices. Then, the \emph{compromised devices detection rate} and \emph{false-positive rate} are defined as
\begin{equation}
\emph{compromised devices detection rate} = \frac{TP}{TP+FN}.
\end{equation}
\begin{equation}
\emph{false-positive rate} = \frac{FP}{TN+FP}.
\end{equation}

\noindent{\textbf{Performance vs. Data used for training}:} We first evaluate the performance against the percentage of data used for training. Fig. \ref{fig:Detection_accuracy_trainrate} shows the compromised devices detection performance against the percentage of joint-FM features or fused FM-TM features used for training. The compromised devices detection rate increases and the false-positive rate decreases as the percentage of data used for training increases. We can also observe that the compromised devices detection rate using fused features is better than using only joint-FM features under the same parameters. The false-positive rate using fused features is lower than using only joint-FM features under the same parameters. Therefore, it is better to combine joint-FM and TM together to accurately distinguish compromised devices from legitimate IoT devices. When 80\% data are used for training, the compromised devices detection rate can reach 88.5\% while the false-positive rate drops under 7.5\%.

\section{Conclusion}\label{sec:conclusion}
In this paper, we propose Linkbait to actively mitigate LFA for IoT by providing a fake linkmap to the adversary. To our knowledge, we are the first to early mitigate LFA before congestion happens, which is totally different from existing works that mitigate LFA after the links are comprised by adversaries. The core of Linkbait is link obfuscation that selectively reroutes probing flows to hide target links from adversaries and mislead them to consider bait links as target links. Furthermore, we extract unique traffic features from both the linkmap construction phase and the flooding phase, and leverage SVM to accurately distinguish compromised devices from legitimate IoT devices. The experiments with real-world testbed and large-scale simulations demonstrate the feasibility and effectiveness of Linkbait. The experimental results show that Linkbait introduces a very small rerouting latency and achieves a high compromised devices detection rate while maintaining a low false positive rate.


\end{document}